\documentclass[runningheads]{llncs}
\pdfoutput=1
\usepackage{llncsdoc}
\let\displaystyle\textstyle

\usepackage{times}
\usepackage{helvet}
\usepackage{courier}
\usepackage{multirow}
\usepackage{url}
\usepackage{balance}

\usepackage{graphicx}

\usepackage[usenames,dvipsnames]{color}
\usepackage{listings}

\newtheorem{algo}{Algorithm}

\newcommand{\ignore}[1]{}

\def\st{\bigskip\noindent}
\newcommand{\lplus}
{
   \stackrel{+}{\gets}
}

\newcommand{\pg}[1]{{\tt #1}}

\def\no{\hbox{\it not\ }}

\newcommand{\horline}{ \rule[1mm] {11.8cm}{0.1mm}}

\newcommand{\ie}[1] {
  \begin{itemize}
    #1
  \end{itemize}
}

\newcommand {\bfigalgo} [2]
{
  \begin{figure}[htb!]
  \begin{tabbing}
    mm\=mm\=mmm\=mmm\=mmm\=mmm\=mmm\=mm\=mm\=mm\=mm\=mm\=mm\=mm \kill \\
    \horline\\
     {\bf #1} #2\\
}
\newcommand {\efigalgo}[1]
{
    \horline
 \end{tabbing}
 \caption{#1}
 \end{figure}
}

\newcommand{\sparc}{${\cal SP \hspace*{-1mm} ARC}$}
\titlerunning{\sparc\ -- Sorted ASP with Consistency Restoring Rules}
\authorrunning{E.~Balai \emph{et al}\/.}

\begin{document}

\setcounter{page}{19}

\title{\sparc\ -- Sorted ASP with Consistency Restoring Rules}

\author{Evgenii Balai\inst{1} \and Michael Gelfond\inst{2} \and Yuanlin Zhang\inst{2}}
\date{}
\institute{Kyrgyz-Russian Slavic University \and Texas Tech University, USA \\ iensen2@mail.ru, michael.gelfond@ttu.edu, y.zhang@ttu.edu}

\maketitle


\begin{abstract}
This is a preliminary report on the work aimed
at making  CR-Prolog -- a version of ASP with consistency
restoring rules -- more suitable for use in teaching and large
applications. First we describe a sorted version
of CR-Prolog  called \sparc.
Second, we translate a basic version of the CR-Prolog into
the language of DLV and compare the performance with the state of the art CR-Prolog solver.
The results form the foundation for future more efficient and user friendly
implementation of  \sparc\ and shed some light on the relationship between two useful knowledge
representation constructs: consistency restoring
rules and weak constraints of DLV.
\end{abstract}

\newcommand{\QED}{$\Box$}

\section{Introduction}
The paper continues work on design and implementation
of knowledge representation languages based on Answer Set
Prolog (ASP) \cite{gl91}.
In particular we concentrate on the extension of ASP
called CR-Prolog -- Answer Set Prolog with consistency
restoring rules (CR-rules for short) \cite{bg03}.
The language, which allows a comparatively simple encoding of indirect
exceptions to defaults, has been successfully used for a number
of applications including planning \cite{Marcello04}, probabilistic
reasoning \cite{gr10}, and reasoning about intentions \cite{jbg}.
This paper is
a preliminary report on our attempts to make CR-Prolog (and hence
other dialects of ASP) more user
friendly and more suitable for use in teaching and large applications.
This work goes in two different, but connected, directions.
First we expand the syntax of CR-Prolog by introducing sorts.
Second, we translate a basic version of the CR-Prolog into the language of DLV
with weak constraints \cite{Faber98} and compare the efficiency of the resulting DLV based CR-Prolog solver with
the CR-Prolog solver implemented in \cite{bald07}. The original hope
for the second part of the work was to
obtain a substantially more efficient inference engine for CR-Prolog.
This was a reasonable expectation -- the older engine is built
on top of existing ASP solvers and hence does not fully
exploit their inner structure.
However this didn't quite work out. Each engine has its strong
and weak points and the matter requires further investigation.
But we believe that even preliminary results are of interest
since they shed some light on the relationship between two useful knowledge
representation constructs: CR-rules and weak constraints.
The first goal requires a lengthier explanation.
Usually, a program of an Answer Set Prolog based language is understood as
a pair, consisting of a signature and a collection of
logic programming rules formed from symbols of this signature.
The syntax of the language does not provide any means
for specifying this signature -- the usual agreement is that the signature
consists of symbols explicitly mentioned in the programming rules.
Even though in many cases this provides a reasonable solution there are also
certain well-known (e.g., \cite{McCainT94,lparse}) drawbacks:
\begin{enumerate}
\item Programs of the language naturally allow unsafe rules which
\begin{itemize}
\item May lead to change of the program behavior under seemingly unrelated
updates. A program $\{p(1). \ \ \ q \leftarrow not~ p(X).\ \ \ \neg q \leftarrow not~q.\}$
entails $\neg q$,
but this conclusion should be withdrawn after adding seemingly unrelated
fact $r(2)$. (This happens because of the introduction of a new
constant $2$ which leads to a new ground instance of the second rule:
$q \leftarrow not~ p(2)$.)
\item Cause difficulties for the implementation of ASP solvers. That is why
most implementations do not allow unsafe rules. The corresponding error messages
however are not always easy to decipher and the elimination of errors is
not always an easy task.
\end{itemize}
\item The language is untyped and therefore  does not provide any protection from
unfortunate typos.
Misspelling $john$ in the fact $parent(jone,mary)$ will not be detected
by a solver and may cost a programmer unnecessary time during the program testing.

\end{enumerate}
There were several attempts to address these problems for ASP
and some of its variants.
The $\#domain$ statements of input language of $lparse$ \cite{lparse} --- a popular grounder
used for a number of ASP systems --- defines sorts for variables.
Even though this device is convenient for simple programs
and allows to avoid repetition of atoms defining sorts of variables
 in the bodies of program's
rules it causes substantial difficulties for medium size and large programs.
It is especially difficult to put together pieces of programs written
at different time or by different people. The same variable may be
declared as ranging
over different sorts by different $\#domain$ statements used in different
programs. So the process of merging these programs
requires renaming of variables.
This concern was addressed by Marcello Balduccini \cite{Marcello07}
whose system, $RSig$, provided
an ASP programmer with means for specifying sorts of parameters of the
language predicates\footnote{In addition, $RSig$ provides simple means
for structuring a program into modules which we will not consider here.}.
$RSig$ is a simple extension of
ASP which does not require any shift in perspective and involves only minor
changes in existing programs. Our new language, \sparc, can be viewed
as a simple modification of $RSig$.
In particular {\em we propose to separate definition of sorts
from the rest of the program and use this separation to improve the type checking
and grounding procedure}.

\section{The Syntax and Semantics of \sparc}
In this section we define a simple variant of
\sparc\ which contains only one predefined sort ${\bf nat}$
of natural numbers. Richer variants may contain other predefined sorts
with precise syntax which would be described in their manuals.
The discussion will be
sufficiently detailed to serve as the basis for the implementation of
\sparc\ reasoning system.

\noindent
Let ${\cal L}$ be a language defined by the following grammar rules:
\begin{verbatim}
<identifier> :- <small_letter> | <identifier><letter> |
        <identifier><digit>
<variable> :- <capital_letter> | <variable><letter> |
        <variable><digit>
<non_zero_digit> :- 1|...|9
<digit> :- 0 | <non_zero_digit>
<positive_integer> :- <non_zero_digit> |
        <positive_integer><digit>
<natural_number> :- 0 | <positive_integer>
<op> :- + | - | * | mod
<arithmetic_term> :- <variable> | <natural_number> |
        <arithmetic_term> <op> <arithmetic_term> |
        (<arithmetic_term>)
<symbolic_function> :- <identifier>
<symbolic_constant> :- <identifier>
<symbolic_term> :- <variable> | <symbolic_constant> |
        <symbolic_function>(<term>,...,<term>)
<term> :- <symbolic_term> | <arithmetic_term>
<arithmetic_rel> :-  = | != | > | >= | < | <=
<pred_symbol> :- <identifier>
<atom> :- <pred_symbol>(<term>,...,<term>) |
        <arithmetic_term> <arithmetic_rel> <arithmetic_term> |
        <symbolic_term> = <symbolic_term> |
        <symbolic_term> != <symbolic_term>
\end{verbatim}
Note  that relations $\mbox{=}$ and $\mbox{!=}$ are defined on pairs of arithmetic and
pairs of non-arithmetic terms. The first is a predefined arithmetic equality,
i.e. $2+3 \mbox{=} 5$, $2+1 \mbox{!=} 1$, etc.
The second is an identity relation\footnote{In the implementation
non-arithmetic identity should be restricted to comply with the syntax of
lparse and other similar grounders.}.
By a {\em ground term} we mean a term containing no variables and no
symbols for arithmetic functions \cite{Gel08}.

\st
From now on we assume a language ${\cal L}$ with a fixed collection
of symbolic constants and predicate symbols.
A \sparc\ program parametrized by ${\cal L}$ consists of
three consecutive parts:

\begin{verbatim}
<program> :-
        <sorts definition>
        <predicates declaration>
        <program rules>
\end{verbatim}

\noindent
{\bf The first part} of the program starts with the keywords:

\st
$sorts\ definition$

\st
and is followed by the sorts definition:
\begin{definition}
{\rm By {\em sort definition} in ${\cal L}$ we mean a collection $\Pi_s$ of
rules of the form
$$a_0 \leftarrow a_{1}, ..., a_{m}, not~ a_{m+1}, ..., not~ a_n.$$
such that
\begin{itemize}
\item $a_i$ are atoms of ${\cal L}$ and $a_0$ contains no arithmetic relations;
\item $\Pi_s$ has a unique answer set ${\cal S}$\footnote{As
usual by ${\cal S}$
we mean answer set of a ground program obtained from $\Pi$ by replacing its
variables with ground terms of ${\cal L}$. We assume that the program has
non-empty Herbrand universe}.
\item For every symbolic ground term $t$ of ${\cal L}$ there is
a unary predicate $s$ such that $s(t) \in {\cal S}$.
\item Every variable occurring in the negative part of the body,
i.e. in at least one of the atoms  $a_{m+1},\dots,a_n$, occurs
in atom $a_i$ for some $0 < i \leq m$.
\end{itemize}
Predicate $s$ such that $s(t) \in {\cal S}$
is called a {\em defined sort} of $t$.  The language can also contain
{\em predefined sorts}, in our case ${\bf nat}$.
Both, defined and predefined sorts will be referred to simply as {\em sorts}.
(Note that a term $t$ may have more than one sort.)
}
\end{definition}
The last condition of the definition is used to
avoid unexpected reaction of the program to introduction
of new constants (see example in the introduction).
The condition was introduced in \cite{McCainT94} where the authors
proved that every program $\Pi$ satisfying this condition has the following
property, called language independence: for every sorted signatures $\Sigma_1$
and $\Sigma_2$ groundings of $\Pi$ with respect to $\Sigma_1$ and
$\Sigma_2$ have the same answer sets. This of course assumes that
every rule of $\Pi$ can be viewed as a rule in $\Sigma_1$ and $\Sigma_2$.

\noindent {\bf The second part} of a \sparc\ program starts with a keyword

\st
$predicates\ declaration$

\st
and is followed by statements of the form

\st
$pred\_symbol(sort,\dots,sort)$

\st
We only allow one declaration per line.
 Predicate symbols occurring in the declaration must differ
from those occurring in sorts definition.
Finally, multiple declarations for one predicate symbol with the same arity are not allowed.

\medskip
\noindent {\bf The third part} of a \sparc\ program
starts with a keyword

\st
$program\ rules$

\st
and is followed by a collection $\Pi_r$ of regular and consistency restoring
rules of \sparc\ defined as follows:

\st
regular rule:
 \begin{equation}
 l_0 \vee\dots\vee l_m \leftarrow l_{m+1},  \ldots, l_k, not~l_{k+1} \ldots not~l_{n}
 \end{equation}
CR-rule:
\begin{equation}
   l_0 \lplus l_1,  \ldots, l_k, not~l_{k+1} \ldots not~l_{n}
 \end{equation}
where $l$'s are literals\footnote{By a literal we mean an atom $a$
or its negation $\neg a$. Note in this paper, we use $\neg$ and \pg{-} interchangeably.} of ${\cal L}$.
Literals occurring in the heads of the rules must not be formed by predicate symbols
occurring in $\Pi_s$. In this paper, $\leftarrow$ and \pg{:-} are used interchangeably, so are $\lplus$ and \pg{:+}.

As expected, program $\Pi_r$ is viewed as a shorthand for the set of all its ground
instances which {\em respect the sorts defined by $\Pi_s$}. Here is the precise
definition of this notion.

\begin{definition}
{\rm
Let $gr(r)$ be a ground instance of a rule $r$ of $\Pi_r$,
i.e. a rule obtained from $r$ by replacing its variables by ground terms of ${\cal L}$.
We'll say that $gr(r)$ respects sorts of $\Pi_s$ if every occurrence
of an atom $p(t_1,\dots,t_n)$ of $gr(r)$
satisfies the following condition:
if $p(s_1,\dots,s_n)$ is the predicate declaration of $p$ then $t_1,\dots,t_n$
are terms of sorts $s_1,\dots,s_n$ respectively.
By $gr(\Pi_r)$ we mean the collection of {\em all}  
ground instances of rules
of $\Pi_r$ which respect sorts of $\Pi_s$.
}
\end{definition}
Note that according to our definition $gr(r)$ may be empty.
This happens, for instance, for a rule which contains
atoms $p_1(X)$ and $p_2(X)$ where $p_1$ and $p_2$ require parameters
from disjoint sorts.



\st
Let us now define answer sets of a ground \sparc\ program $\Pi$.
We assume that the readers are familiar with the definition of answer sets
for standard ASP programs.
Readers unfamiliar with the intuition behind the notion of consistency restoring rules
of CR-Prolog are referred to the Appendix.


\st
First we will need some notation.
The set  of regular rules of a \sparc\ program $\Pi$ will be denoted by
$R$; the set of cr-rules of $\Pi$ will be denoted by $CR$.
By $\alpha(r)$ we denote a regular rule
obtained from a consistency restoring rule $r$
by replacing $\lplus$ by $\leftarrow$;
$\alpha$ is expanded in a standard way to a set $X$ of cr-rules,
i.e. $\alpha(X) = \{\alpha(r)\; :\; r \in X\}$.

\begin{definition}{(Abductive Support)}\\
{\rm A 
collection $X$ of cr-rules of $\Pi$ such that
\begin{enumerate}
\item $R \cup \alpha(X)$ is consistent (i.e. has an answer set) and
\item any $R_0$ satisfying the above condition has cardinality
which is greater than or equal to that of $R$
\end{enumerate}
is called an {\em abductive support} of $\Pi$.
}
\end{definition}
\begin{definition}{(Answer Sets of \sparc\ Programs)}\\
{\rm A set $A$ is called an \emph{answer set} of $\Pi$ if it is an answer set
of a regular program $R \cup \alpha(X)$ for some abductive
support $X$ of $\Pi$.}
\end{definition}

 To complete the definition of syntax and semantics of a \sparc\
program we need to note that though such program is defined
with respect to some language ${\cal L}$ in practice this
language is extracted from the program. We always assume
that terms of ${\cal L}$ defined by a \sparc\ program $P$
are arithmetic terms and terms defined
by the sorts definition\footnote{A term $t$ is defined by $\Pi_s$ if for some sort $s$, $s(t)$ belongs to the answer set of $\Pi_s$}; predicate symbols are those occurring in
sorts definition and predicate declaration.
Now we are ready to give an example of a \sparc\ program.

\begin{example}\label{ex1}[\sparc\ programs]\\
Consider a \sparc\  program $P_1$:
\begin{verbatim}
sorts definition
s1(1).
s1(2).
s2(X+1) :-
         s1(X).
s3(f(X,Y)) :-
         s1(X),
         s1(Y),
         X != Y.
predicates declaration
p(s1)
q(s1,s3)
r(s1,s3)
program rules
p(X).
r(1,f(1,2)).
q(X,Y) :-
       p(X),
       r(X,Y).
\end{verbatim}

 The sort declaration of the program defines ground
terms $1$, $2$, $3$, $f(1,2)$, $f(2,1)$
with the following defined sorts:

\st
$s_1 = \{1,2\}$\\
$s_2 = \{2,3\}$\\
$s_3 = \{f(1,2),f(2,1)\}$

\st  Of course, $1$, $2$, and $3$ are also of the sort ${\bf nat}$.
The sort respecting grounding of the rules of $\Pi$ is
\begin{verbatim}
p(1).
p(2).
r(1,f(1,2)).
q(1,f(1,2)) :-
       p(1),
       r(1,f(1,2)).
q(2,f(1,2)) :-
       p(2),
       r(2,f(1,2)).
q(1,f(2,1)) :-
       p(1),
       r(1,f(2,1)).
q(2,f(2,1)) :-
       p(2),
       r(2,f(2,1)).
\end{verbatim}

\st
The answer set of the program is $\{p(1),p(2),r(1,f(1,2)),q(1,f(1,2))\}$.
(We are not showing the sort atoms.)

\medskip
Consider now a \sparc\  program $P_2$:

\begin{verbatim}
sorts definition
t(a,b).
t(c,1).
s1(X) :- t(X,Y).
s2(Y) :- t(X,Y).
s3(a).
predicates declaration
p(s1,s2).
program rules
p(X,Y) :- s3(X),t(X,Y).
\end{verbatim}
The sort respecting grounding of the program is
\begin{verbatim}
p(a,b) :- s3(a),t(a,b).
\end{verbatim}
Its answer set is $\{p(a,b),t(a,b)\}$.

\medskip
Another example can be obtained by restating the CR-Prolog  program
from Example \ref{e2} in the Appendix by adding sort definitions
$s_1(a)$ and $s_2(d(a))$ and predicates declarations $p(s_1)$, $q(s_1)$,
$c(s_1)$ and $ab(s_2)$. One can easily check that, as expected,
the answer set of the resulting program is
$\{\neg q(a),c(a),\neg p(a))\}$.
\end{example}

\section{Translation of \sparc\ Programs to DLV Programs}

DLV \cite{dlv06} is one of the well developed solvers for ASP
programs. We select DLV as the target language mainly because of
its {\em weak constraints} \cite{Faber98} which can be used to
represent cr-rules. A {\em weak constraint} is of the form
$$:\sim l_1, \ldots, l_k, not~l_{k+1} \ldots not~l_{n}.$$
where $l_i$'s are literals. (Weak constraints of DLV allow
preferences which we ignore here.)
Informally, weak constraints can be violated, but as many of them
should be satisfied as possible. The {\em answer sets} of a program
$P$ with a set $W$ of weak constraints are those of $P$ which
minimize the number of violated weak constraints.

\st We first introduce some notations before presenting the translation algorithm.

\begin{definition}{(DLV counterparts of \sparc\ programs)}\\
{\rm A DLV program $P_2$ is {\em a counterpart} of \sparc\ program $P_1$
if answer sets of $P_1$ and $P_2$ coincide on literals from the
language of $P_1$.}
\end{definition}

\begin{definition}
{\rm
Given a \sparc\ program $P$, we associate a unique number
to each of its cr-rules.
The {\em name of a cr-rule} $r$ of $\Pi$ is a term
$rn(i, X_1, ..., X_n)$ where $rn$ is a new function symbol,
$i$ is the unique number associated with $r$,
and $X_1, ..., X_n$ is the list of distinct variables occurring in $r$.}
\end{definition}
 For instance, if rule $p(X,Y) \leftarrow q(Z,X,Y)$ is assigned number $1$ then
its name is $rn(1,X,Y,Z)$.

\noindent In what follows we describe a translation of \sparc\ programs into their DLV counterparts.

\begin{algo}\label{transl} (\sparc\ program translation)

{\bf Input}: a \sparc\ program $P_1$.

{\bf Output}: a DLV counterpart $P_2$ of $P_1$.
\ie{
  {\em
    \item[1.]  Set variable $P_2$ to $\emptyset$, and let \pg{appl}/1 be a new predicate not occurring in $P_1$.
    \item[2.]  Add all rules of the sorts definition part of $P_1$ to $P_2$.
    \item[3.]  For any program rule $r$ of $P_1$,
    \ie{
    \item [3.1.] Let
$$s=\{s_1(t_1), ..., s_n(t_n)\ |\ p(t_1, ..., t_n) \mbox{
occurs in } r \mbox{ and  } p(s_1, ...,s_n) \in P_1 \},$$
and let rule $r^\prime$ be the result of adding all elements of $s$ to the body of $r$.
    \item[3.2.]  If $r^\prime$ is a regular rule, add it to $P_2$.
    \item[3.3.]  If $r^\prime$ is a cr-rule of the form

    $$q \lplus body.$$

     add to $P_2$ the rules

    {\tt
    \begin{tabbing}
    appl($rn(i, X_1, ..., X_n)$)$\lor$ $\neg$appl($rn(i, X_1, ..., X_n)$) :- $body$. \\
    :$\sim$ appl($rn(i, X_1, ..., X_n)$), $body$. \\
    $q$ :- appl($rn(i, X_1, ..., X_n)$), $body$.
    \end{tabbing}
    }
where $rn(i,X_1, ..., X_n)$ is the name of $r$.}
  }
}
\end{algo}

\noindent The intuitive idea behind the rules added to $P_2$ in 3.3. is
as follows:
\pg{appl($rn(i,X_1,$} $..., X_n)$) holds if the cr-rule $r$ is used to
obtain an answer
set of the \sparc\ program; the first rule says that $r$
is either used or not used; the second
rule, a weak constraint, guarantees that
$r$ is not used if possible,
and the last
rule allows the use of $r$ when necessary.

\st The correctness of the algorithm
is guaranteed by the following theorem whose complete proof can be found at http://www.cs.ttu.edu/research/krlab/pdfs/papers/sparc-proof.pdf.
\begin{theorem}\label{th1}
A DLV program $P_2$ obtained from a \sparc\ program $P_1$ by
Algorithm 1 is a DLV counterpart of $P_1$.
\end{theorem}

\noindent The translation can be used to compute an answer set
of \sparc\ program $P$.

\begin{algo}\label{transl} (Computing an answer set of a \sparc\ program)

{\bf Input}: a \sparc\ program $P$.

{\bf Output}: an answer set of $P$.
\ie{ {\em
\item[1] Translate $P$ into its DLV counterpart $P^\prime$.
\item[2] Use DLV to find an answer set $S$ of $P^\prime$.
\item[3] Drop all literals with predicate symbol \pg{appl} from $S$ and return the new set.
}}
\end{algo}

\begin{example}
\noindent To illustrate the translation and the algorithm, consider the following program.

{\tt
\begin{tabbing}
sorts definition \\
s(a). \\
predicates declaration \\
p(s) \\
q(s)  \\
program rules  \\
p(X) :- not q(X).  \\
-p(X).   \\
q(X) :+ .
\end{tabbing}
}

\noindent After step 2 of Algorithm 1 , $P^\prime$ becomes:

{\tt
\begin{tabbing}
 s(a).
\end{tabbing}
}

\noindent After the execution of the loop 3 of this algorithm
for the first and second program rule, $P^\prime$ becomes
{\tt
\begin{tabbing}
 s(a). \\
p(X) :- not q(X),s(X). \\
$\neg$p(X):- s(X).
\end{tabbing}
}

\noindent Assuming the only cr-rule is numbered by 1, after the algorithm is applied to
the third rule, $P^\prime$ becomes
{\tt
\begin{tabbing}
 s(a). \\
p(X) :- not q(X),s(X). \\
$\neg$p(X):- s(X). \\
appl($rn(1,X)$) $\lor$ $\neg$appl($rn(1,X)$) :- s(X). \\
:$\sim$~appl($rn(1,X)$), s(X). \\
q(X) :- appl($rn(1, X)$), s(X).
\end{tabbing}
}

\noindent Given the program $P^\prime$, DLV solver
returns an answer set
$$\{s(a), appl(rn(1, a)),  q(a),  \neg p(a)\}$$
After dropping \pg{appl($rn(1, a)$)} from this answer set, we obtain an
answer set
$$\{s(a), q(a), \neg p(a)\}$$
for the original program.
\end{example}

\section{Experimental Results}
We have implemented a \sparc\ program solver, called {\em crTranslator} (available from the link in \cite{balai12}), based on the proposed translation approach. CRModels2 \cite{bald07} is the state of the art solver for CR-prolog programs. To compare the performance of the DLV based solver to CRModels2, we use the classical benchmark of the reaction control system for the space shuttle \cite{Marcello04} and new benchmarks such as representing and reasoning with intentions \cite{jbg}, and the shortest path problem.

\st Clock time, in seconds, is used to measure the performance of the solvers. Since the time complexity of translation is low, the recorded problem solving time does not include the translation time.

\st In this experiment, we use DLV build BEN/Dec 21 2011 and CRModels2 2.0.12 \cite{Bal12} which uses ASP solver Clasp 2.0.5 with grounder Gringo 3.0.4 \cite{clasp07}. The experiments are carried out on a computer with Intel Core 2 Duo CPU E4600 at 2.40 Ghz, 3GB RAM, and Cygwin 1.7.10 on Windows XP.

\subsection{The First Benchmark: Programs for Representing and Reasoning with Intentions}
Recently, CR-Prolog has been employed to represent and reason with intentions \cite{jbg}. We compare crTranslator with CRModels2 on the following scenarios proposed in \cite{jbg}:
Consider a row of four rooms, $r_1,r_2,r_3,r_4$ connected by doorways, such that an
agent may move along the row from one room to the next. We say that two
people {\em meet} if they are located in the same room. Assume that initially our
agent Bob is in $r_1$ and he intends to meet with John who, as Bob knows, is
in $r_3$. This type of intention is frequently referred to as an intention to achieve
the goal. The first task is to design a simple plan for Bob to
achieve this goal: move from $r_1$ to $r_2$ and then to $r_3$. Assuming that as Bob is moving from $r_1$ to $r_2$, John moves from $r_3$ to $r_2$, the second task is to recognize the unexpected achievement of his goal and not continue moving to $r_3$. Programs to implement these two tasks are given as $\mathcal{B}_0$ and $\mathcal{B}_1$ respectively in \cite{jbg}.

\begin{figure}
\begin{center}
\begin{tabular}{|c|c|c|} \hline
    Tasks  & CRModels2 & crTranslator\\ \hline
    task 1 & 104 & {\bf 11}   \\ \hline
    task 2 & 104 & {\bf 101} \\ \hline
\end{tabular}
\end{center}
\caption{\label {fig:intention} CPU time for intention reasoning benchmark using CRModels2 and crTranslator}
\end{figure}

\noindent In this experiment, crTranslator has a clear advantage over CR-Models2 on task 1 and similar performance on task 2.

\subsection{The Second Benchmark: Reaction Control System of Space Shuttle}
USA-Smart is a CR-prolog program to find plans with improved quality for the operation of the Reaction Control System (RCS) of the Space Shuttle. Plans consist of a sequence of operations to open
and close the valves controlling the flow of propellant from the tanks to the jets of the RCS.

\st In our experiment, we used the USA-Smart program with four instances: fmc1 to fmc4 \cite{usaSmart}. The \sparc\ variant of the USA-Smart program is written as close as possible to USA-smart.
The results of the performance of crTranslator and CRModels for these programs are listed in Figure~\ref{fig:usaSmart}.

\begin{figure}
\begin{center}
\begin{tabular}{|c|c|c|c|c|c|} \hline
    Instances & CRModels2 & crTranslator\\ \hline
    fmc1 & {\bf 29.0} &  74.0  \\ \hline
    fmc2 & {\bf 11.6} & 34.0    \\ \hline
    fmc3 & {\bf 6.0} & 8907.0  \\ \hline
    fmc4 & {\bf 30.5} & 22790.0 \\ \hline
\end{tabular}
\end{center}
\caption{\label {fig:usaSmart} CPU time for reaction control system using CRModels2 and crTranslator}
\end{figure}

 \st We note that these instances have small abductive supports (with sizes of the supports less than 9) and relatively large number of cr-rules (with more than 1200).
This can partially explain why CRModels2 is faster because it finds the abductive support by exhaustive enumeration of the candidate supports starting from size 0 to all cr-rules in an increasing manner.

\subsection{The Third Benchmark: Shortest Path Problem} Given a simple directed graph and a pair of distinct vertices of the graph, the shortest path problem is to find a shortest path between these two vertices. Given a graph with $n$ vertices and $e$ edges, its {\em density} is defined as $e/(n * (n-1))$. In our experiment, the problem instances are generated randomly based on the number of vertices and the density of the graph. The density of the graphs varies from 0.1 to 1 so that the shortest paths involve abductive supports of different sizes.
To produce graphs with longer shortest path (which needs larger abductive supports), we zoom into the density between 0 to 0.1 with a step of 0.01. To reduce the time solving the problem instances, as density increases, we use smaller number of vertices. Given a graph, we define {\em the distance} between a pair of vertices as the length of the shortest path between them. For any randomly generated graph, we select any two vertices such that their distance is the longest among those of all pairs of vertices. The problem is to find the shortest path between these two vertices.

\st The \sparc\ programs and CR-prolog programs are written separately due to the difference between these two languages, but we make them as similar as possible and use exactly the same cr-rules in both programs. The experimental results are listed in Figure~\ref{fig:graph}.

\st From the results, CRModels2 is faster on a majority of cases. Again, crTranslator is faster when the size of the abductive support is large. The graphs with density between 0.02 and 0.03 have support size of 16 while the other graphs (except the one of density 0.01) have support sizes not more than 12. Further investigation is needed to have a better understanding of the performance difference between the two solvers.
\begin{figure}
\begin{center}
\begin{tabular}{|c|c|c|c|c|c|} \hline
Number of vertices & Density & CRModels2& crTranslator \\ \hline
60&0.01&4.0&{\bf 0.2}  \\ \hline
60&0.02&5.1&{\bf 0.4} \\ \hline
60&0.03&5.7&{\bf 0.8} \\ \hline
60&0.04&{\bf 9.1}&66.5 \\ \hline
60&0.05&{\bf 29.2}&337.0 \\ \hline
60&0.06&{\bf 235.7}& 4451.8\\ \hline
40&0.07&{\bf 7.4}&19.9 \\ \hline
40&0.08&{\bf 8.4}&154.6 \\ \hline
40&0.09&{\bf 7.0}&32.6\\ \hline
30&0.1&{\bf 6.0}&16.8 \\ \hline
30&0.2&{\bf 39.9}&9711.4\\ \hline
20&0.3&{\bf 7.6}&54.9 \\ \hline
20&0.4&{\bf 9.3}&52.2 \\ \hline
20&0.5&{\bf 16.4}&234.8 \\ \hline
20&0.6&{\bf 9.6}&51.7  \\ \hline
20&0.7&{\bf 14.3}&52.0\\ \hline
20&0.8&{\bf 17.6}&58.8\\ \hline
20&0.9&{\bf 22.2}&69.1 \\ \hline
20&1.0&{\bf 5.5}&55.6\\ \hline
\end{tabular}
\end{center}
\caption{\label{fig:graph} CPU time for solving shortest path problem using CRModels2 and crTranslator}
\end{figure}

\section{Conclusion and Future Work}
This paper describes a sorted version
of CR-Prolog  called \sparc,
presents a translation of consistency restoring rules
of the language into weak constraints of DLV,
and investigates the possibility of building efficient
inference engines for \sparc\ based on this translation.
This is a preliminary report. There is a number of
steps which should be made to truly develop \sparc\
into a knowledge representation language of choice
for teaching and applications. In particular we plan
the following:

\begin{itemize}
\item Expand \sparc\ to include a number of useful language
constructs beyond the original
language of ASP such as aggregates and optimization constructs.
In this expansion, instead of committing to a particular syntax,
we are planning to allow users to select
their favorite input language such as that of DLV or LPARSE or GRINGO
and provide the final system with support for the corresponding language.

\item Provide \sparc\ with means to specify different preference
relations between sets of cr-rules, define and investigate
answer sets minimal with respect to these preference relations,
and implement the corresponding \sparc\ solvers.

\item Design and implement \sparc\ grounders to directly
use the sort information provided by definitions and declaration
of a program. The emphasis will be on error checking and incrementality
of the grounders.

\item Investigate more efficient reasoning algorithms for \sparc. DLV uses a more advanced technique of branch and bound to process weak constraints while CRModels employs a more primitive search algorithm. However, our experiments show that the latter is not necessarily slower. Further understanding of these two approaches is expected to inspire new techniques for building more efficient solvers for \sparc\ programs.

\item Expand \sparc\ and its solvers to other extensions
of ASP including ACC \cite{MellarkodGZ08} and P-log \cite{gr10}.

\end{itemize}

\section*{Acknowledgement}
The work of Gelfond and Zhang was partially supported by NSF grant IIS-1018031.

\appendix
\section*{Appendix: CR-Prolog}
This Appendix contains a short informal introduction to CR-Prolog.
The version discussed here is less general than the standard version ---
in particular it omits the treatment of preferences which is a task
orthogonal to the goals of this paper.
One of the original goals of the CR-Prolog was to provide a construct
allowing a simple representation of exceptions to
defaults, sometimes referred to as
{\bf indirect exceptions}.
Intuitively,
these are rare exceptions that come into play only as a last resort, to
restore the consistency of the agent's world view when all else fails.
The representation of indirect exceptions
seems to be beyond the power of  ``pure'' ASP \cite{gl91} which prompted the introduction of cr-rules.
To illustrate the problem let us consider the following example.

\begin{example}\label{e1}[Indirect Exception in ASP]\\
Consider an ASP representation of the default ``elements of class
$c$ normally have property $p$'':
\[
p(X) \leftarrow c(X), \no ab(d(X)), \no \neg p(X).
\]
(where $d(X)$ is used as the name of the default)
together with the rule
\[
q(X) \leftarrow  p(X).
\]
and two observations:
\[
\begin{array}{c}
\ \ c(a). \\
\neg q(a).
\end{array}
\]

\noindent It is not difficult to check that this program is inconsistent.
No rules allow the reasoner to prove that
the default is not applicable to $a$ (i.e. to prove $ab(d(a))$) or that
$a$ does not have property $p$. Hence the default must conclude
$p(a)$. The second rule implies $q(a)$ which contradicts the observation.

\st There, however, seems to exist a commonsense argument which
may allow a reasoner to avoid inconsistency, and to conclude
that  $a$ is an indirect exception to the default.
The argument is based on the
{\bf Contingency Axiom}
for default
$d(X)$ which says that
{\em Any element of class $c$ can be an exception to the default $d(X)$ above,
but such a possibility
is very rare and, whenever possible, should be ignored}.
One may informally argue that since the application of the default to $a$
leads to a contradiction, the possibility of $x$ being an exception
to $d(a)$ cannot be ignored and hence
$a$ must satisfy this rare property.
\end{example}

\st The CR-Prolog is an extension of ASP capable of encoding
and reasoning about such rare events.
In addition to regular logic programming rules the language allows
{\em consistency restoring} rules of the form
\begin{equation}\label{cr-rules}
l_0 \lplus  l_{1}, \ldots, l_k, \no l_{k+1},\ldots,\no l_n
\end{equation}
where $l$'s are literals.
Intuitively, the rule says that if the reasoner
associated with the program believes the body of the rule, then it
``may possibly'' believe its head.
However, this possibility may be used only if there is no way
to obtain a consistent set of beliefs by using only regular rules of the
program.

\st The following Example shows the use of CR-Prolog for representing defaults
and their indirect exceptions.
\begin{example}\label{e2}[Indirect Exception in CR-Prolog]\\
The CR-Prolog representation of default $d(X)$ may look as follows
\[
\begin{array}{c}
p(X) \leftarrow c(X), \no ab(d(X)), \no \neg p(X). \\
\neg p(X)\lplus  c(X).
\end{array}
\]
The first rule is the standard ASP representation of the default,
while the second rule expresses the Contingency Axiom for default
$d(X)$. Consider now a program obtained by combining these two rules
with an atom $c(a).$

\noindent Assuming that $a$ is the only constant in the signature of this program,
the program's answer set will be $\{c(a), p(a)\}$.
Of course this is also the answer set of the regular part of our program.
(Since the regular part is consistent, the Contingency Axiom is ignored.)
Let us now expand this program by the rules
\[
\begin{array}{c}
q(X) \leftarrow p(X). \\
\neg q(a).
\end{array}
\]

\noindent The regular part of the new program is inconsistent. To save the day
we need to use the Contingency Axiom for $d(a)$ to form the abductive support
of the program. As a result
the new program has the answer set $\{\neg q(a),c(a),\neg p(a))\}$.
The new information does not produce inconsistency as in the
analogous case of ASP representation. Instead the program
withdraws its previous conclusion and recognizes $a$ as a (strong) exception
to default $d(a)$.
\end{example}

\noindent The possibility to encode rare events which may serve as unknown exceptions
to defaults proved to be very useful for various knowledge representation
tasks, including planning, diagnostics, and reasoning about the agent's
intentions.

\bibliography{biblio}
\bibliographystyle{splncs}

\end{document}